\documentclass[
 reprint,
 amsmath,amssymb,
 aps,nofootinbib,superscriptaddress
]{revtex4-2}

\usepackage{graphicx}
\usepackage{dcolumn}
\usepackage{bm}
\usepackage{color}

\usepackage{mathtools}
\usepackage{physics}
\usepackage{gensymb}
\usepackage{amsmath, amssymb, amsthm}
\usepackage{epstopdf}

\begin{document}

\preprint{APS/123-QED}

\title{Hierarchy of Lifshitz transitions in the surface electronic structure of\\ Sr\textsubscript{2}RuO\textsubscript{4} under uniaxial compression}

\author{Edgar Abarca Morales}
\affiliation{Max Planck Institute for Chemical Physics of Solids, N\"othnitzer Strasse 40, 01187 Dresden, Germany}
\affiliation{SUPA, School of Physics and Astronomy, University of St Andrews, St Andrews KY16 9SS, UK}

\author{Gesa-R. Siemann}
\affiliation{SUPA, School of Physics and Astronomy, University of St Andrews, St Andrews KY16 9SS, UK}

\author{Andela Zivanovic}
\affiliation{Max Planck Institute for Chemical Physics of Solids, N\"othnitzer Strasse 40, 01187 Dresden, Germany}
\affiliation{SUPA, School of Physics and Astronomy, University of St Andrews, St Andrews KY16 9SS, UK}

\author{Philip A. E. Murgatroyd}
\affiliation{SUPA, School of Physics and Astronomy, University of St Andrews, St Andrews KY16 9SS, UK}

\author{Igor Markovi{\'c}}
\altaffiliation{Present Address: Quantum Matter Institute, University of British Columbia, Vancouver V6T 1Z4, BC, Canada}
\affiliation{Max Planck Institute for Chemical Physics of Solids, N\"othnitzer Strasse 40, 01187 Dresden, Germany}
\affiliation{SUPA, School of Physics and Astronomy, University of St Andrews, St Andrews KY16 9SS, UK}

\author{Brendan~Edwards}
\affiliation{SUPA, School of Physics and Astronomy, University of St Andrews, St Andrews KY16 9SS, UK}

\author{Chris A. Hooley}
\affiliation{SUPA, School of Physics and Astronomy, University of St Andrews, St Andrews KY16 9SS, UK}

\author{Dmitry A. Sokolov}
\affiliation{Max Planck Institute for Chemical Physics of Solids, N\"othnitzer Strasse 40, 01187 Dresden, Germany}

\author{Naoki Kikugawa}
\affiliation{National Institute for Materials Science, Tsukuba, Ibaraki 305-0003, Japan}

\author{Cephise Cacho}
\author{Matthew D. Watson}
\author{Timur K. Kim}
\affiliation{Diamond Light Source, Harwell Science and Innovation Campus, Didcot, OX11 ODE, United Kingdom}

\author{Clifford W. Hicks}
\affiliation{Max Planck Institute for Chemical Physics of Solids, N\"othnitzer Strasse 40, 01187 Dresden, Germany}
\affiliation{School of Physics and Astronomy, University of Birmingham, Birmingham B15 2TT, UK}

\author{Andrew P. Mackenzie}
\affiliation{Max Planck Institute for Chemical Physics of Solids, N\"othnitzer Strasse 40, 01187 Dresden, Germany}
\affiliation{SUPA, School of Physics and Astronomy, University of St Andrews, St Andrews KY16 9SS, UK}

\author{Phil D. C. King}
\email{pdk6@st-andrews.ac.uk}
\affiliation{SUPA, School of Physics and Astronomy, University of St Andrews, St Andrews KY16 9SS, UK}

\date{\today}

\begin{abstract}
We report the evolution of the electronic structure at the surface of the layered perovskite Sr$_2$RuO$_4$ under large in-plane uniaxial compression, leading to anisotropic $B_{1g}$ strains of ${\varepsilon_{xx}-\varepsilon_{yy}=-0.9\pm0.1\%}$. From angle-resolved photoemission, we show how this drives a sequence of Lifshitz transitions,  reshaping the low-energy electronic structure and the rich spectrum of van Hove singularities that the surface layer of Sr$_2$RuO$_4$ hosts. From comparison to tight-binding modelling, we find that the strain is accommodated predominantly by bond-length changes rather than modifications of octahedral tilt and rotation angles. Our study sheds new light on the nature of structural distortions at oxide surfaces, and how targeted control of these can be used to tune density of states singularities to the Fermi level, in turn paving the way to the possible realisation of rich collective states at the Sr$_2$RuO$_4$ surface.  
\end{abstract}

\maketitle

A central building block of numerous correlated electron materials is the transition-metal-oxide octahedron. The distortions of coupled octahedra away from idealised cubic geometries underpin many of the striking physical properties which transition-metal oxides host. In perovskite nickelates, for example, tilts and rotations combined with breathing-like distortions of the NiO$_6$ octahedra support a rich phase diagram of metal-insulator and magnetic transitions~\cite{Torrance1992,Varignon2017}; in several titanates, off-centering of the Ti atom within the octahedral cage generates a ferroelectric state~\cite{Takahasi2013,Tsuda2012a}; while in some manganites, tri-linear coupling of non-polar tilt and rotation modes with polar displacements creates novel multiferroics~\cite{Benedek2011}. In the ruthenate family, modest structural distortions drive the emergence of numerous correlated electron states~\cite{Dagotto2005,Carlo2012,Ortmann2013,Lu2013,Herklotz2016}: unconventional superconductors~\cite{Mackenzie2017a}, Mott insulators~\cite{Nakatsuji2013}, polar metals~\cite{Lei2018}, and quantum criticality~\cite{Grigera2001} are all found in systems built around nominally the same RuO$_6$ structural unit. Disentangling the structure-property relations underpinning the formation of such disparate ground states is a major challenge in the field. 

To this end, developing routes to observe modifications in electronic properties when structural distortions are tuned in a controlled manner is a key goal. Uniaxial pressure can provide such a control parameter~\cite{Ikeda2004,Ikeda2013,Hicks2014b,Kim2018}, and can be applied in conjunction with spectroscopic probes~\cite{Sunko2019b,Ricco2018a,Flototto2018,Lin2021,King2021,Potapenko2014,Yim2018}. In Sr$_2$RuO$_4$, for example, uniaxial compression has been shown to more than double its superconducting $T_\mathrm{c}$ and to stabilise $T$-linear resistivity~\cite{Steppke2017a,Barber2018b}. Both effects have been attributed to a strain-driven Lifshitz transition in the electronic structure, where a saddle point van Hove singularity (vHS), and its associated peak in the density of states, is driven through the Fermi level~\cite{Barber2018b,Sunko2019b}.

\begin{figure*}
\includegraphics{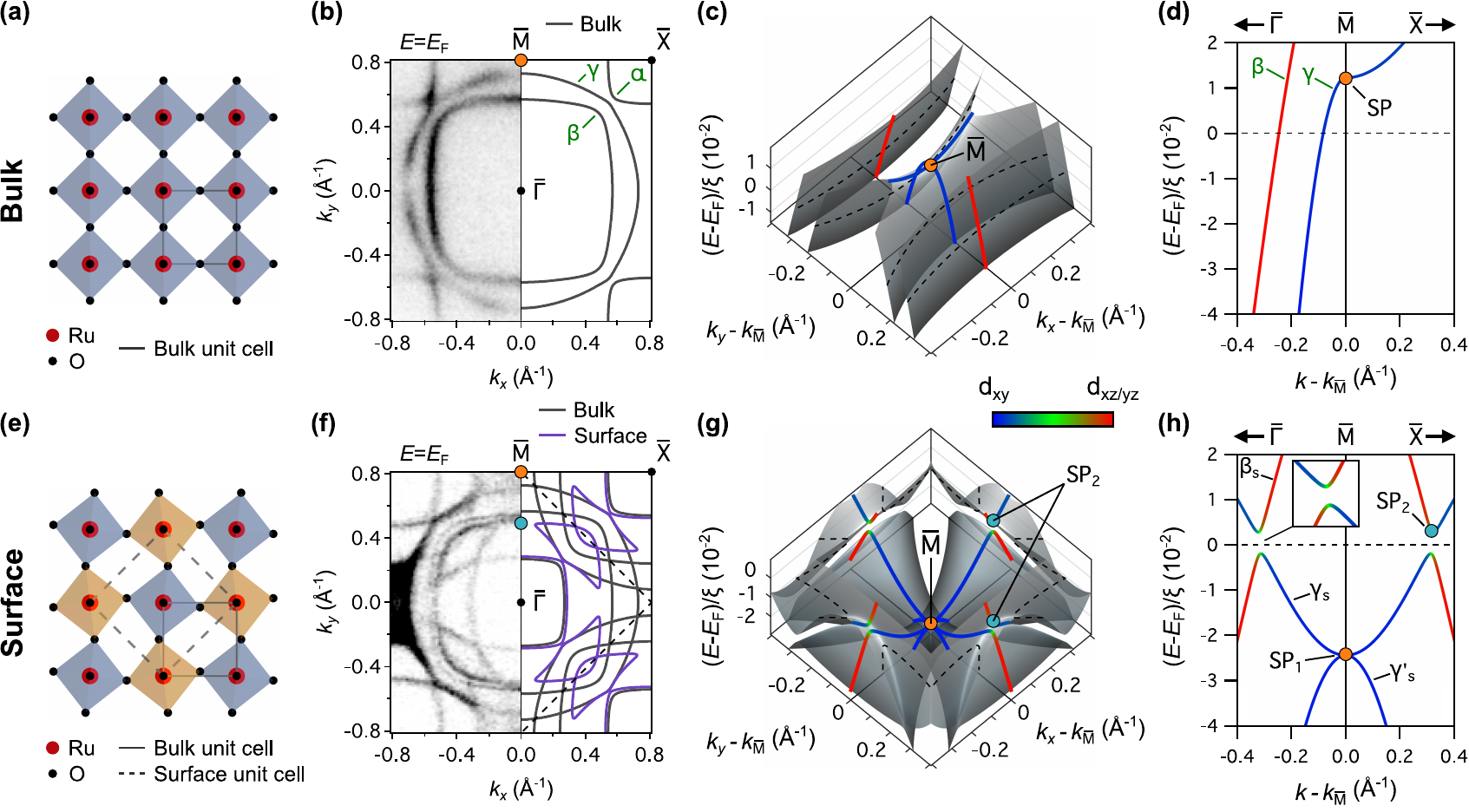}
\caption{\label{fig:fig1} (a) Top view of the RuO\textsubscript{2} layer of bulk Sr\textsubscript{2}RuO\textsubscript{4}. (b) Bulk Fermi surface measured using ARPES (left, reproduced from Ref.~\cite{Sunko2019b}) and calculated from our tight binding model (right). (c) Calculated electronic structure in the vicinity of the $\overline{\mathrm{M}}$-point, showing the bulk vHS arising from the saddle point (SP) of the $\gamma$-band. (d) Corresponding calculated dispersions along $\overline{\mathrm{\Gamma}}$-$\overline{\mathrm{M}}$-$\overline{\mathrm{X}}$. (e) Bipartite RuO\textsubscript{2} layer of the surface of Sr\textsubscript{2}RuO\textsubscript{4}. (f) Surface Fermi surface measured with ARPES (left, $h\nu$=100eV, LV-pol.) and calculated from a tight-binding model including the octahedral rotation (right). (g,h) Corresponding calculated electronic structure of the surface bands in the vicinity of the $\overline{\mathrm{M}}$-point. $\xi$ is the bandwidth of the unstrained surface electronic structure (see Supplemental Fig.~S4~\cite{SM}).}
\end{figure*}

Here we report the observation, from angle-resolved photoemission (ARPES), of the influence of uniaxial pressure on the surface electronic structure of Sr$_2$RuO$_4$. The Sr$_2$RuO$_4$ surface is known to distort via in-plane rotations of its RuO$_6$ octahedra, forming distinct electronic states with significantly more complex Fermi surfaces and low-energy electronic structures as compared to the bulk (Fig.~\ref{fig:fig1})~\cite{Damascelli2000}. It thus serves as a benchmark system for probing the influence of small structural distortions on the electronic states. Our measurements and comparison with model calculations allow us to track how these are modified with strain. Through this, we show that bond-length distortions, not additional octahedral rotations, dominate the strain response in the surface layer, in turn mediating a rich sequence of surface Lifshitz transitions.

High-resolution ARPES measurements were performed using the I05 beamline at Diamond Light Source. Single-crystal samples were grown by the floating zone method \cite{Bobowski2019}. Unlike in Ref.~\cite{Sunko2019b}, where the samples were cleaved {\it ex situ} to remove signatures of surface states, here we cleave {\it in situ} at the measurement temperature of $\approx7$~K. This produces a clean surface with a well-ordered $\sqrt{2}\times\sqrt{2}$ reconstruction. Strain was applied through differential thermal contraction, using a compact, bimetallic platform described in Ref. \cite{Sunko2019b} (see also Supplementary Fig.~S1(a-c)~\cite{SM}). The induced anisotropic sample strain was characterized optically as shown in Supplemental Fig.~S1(d)~\cite{SM}. 

Sr$_2$RuO$_4$ is comprised of single layers of corner-sharing RuO$_6$ octahedra (Fig.~\ref{fig:fig1}(a)), separated by SrO rocksalt layers. The conducting RuO$_2$ layers yield a quasi-two-dimensional three-band Fermi surface with states derived from the three partially-occupied $t_{2g}$ orbitals (Fig.~\ref{fig:fig1}(b))~\cite{Mackenzie1996b}. In the surface layer, the RuO$_6$ octahedra are rotated about the $c$-axis by $\approx6-10^\circ$~\cite{Matzdorf2002}, in anti-phase on neighboring sites (Fig.~\ref{fig:fig1}(e)) creating a 2 Ru-atom unit cell. The bulk states become backfolded about the new Brillouin zone boundary, while additional surface states are split off from the bulk manifold (Fig.~\ref{fig:fig1}(f))~\cite{Damascelli2000,Veenstra2013}. Both the bulk and surface fermiology are well described by a simple tight-binding model, as shown in Fig.~\ref{fig:fig1}(b,f) and discussed in more detail in the Supplemental Material~\cite{SM} (Figs.~S2-S6).

\begin{figure}[h]
\includegraphics{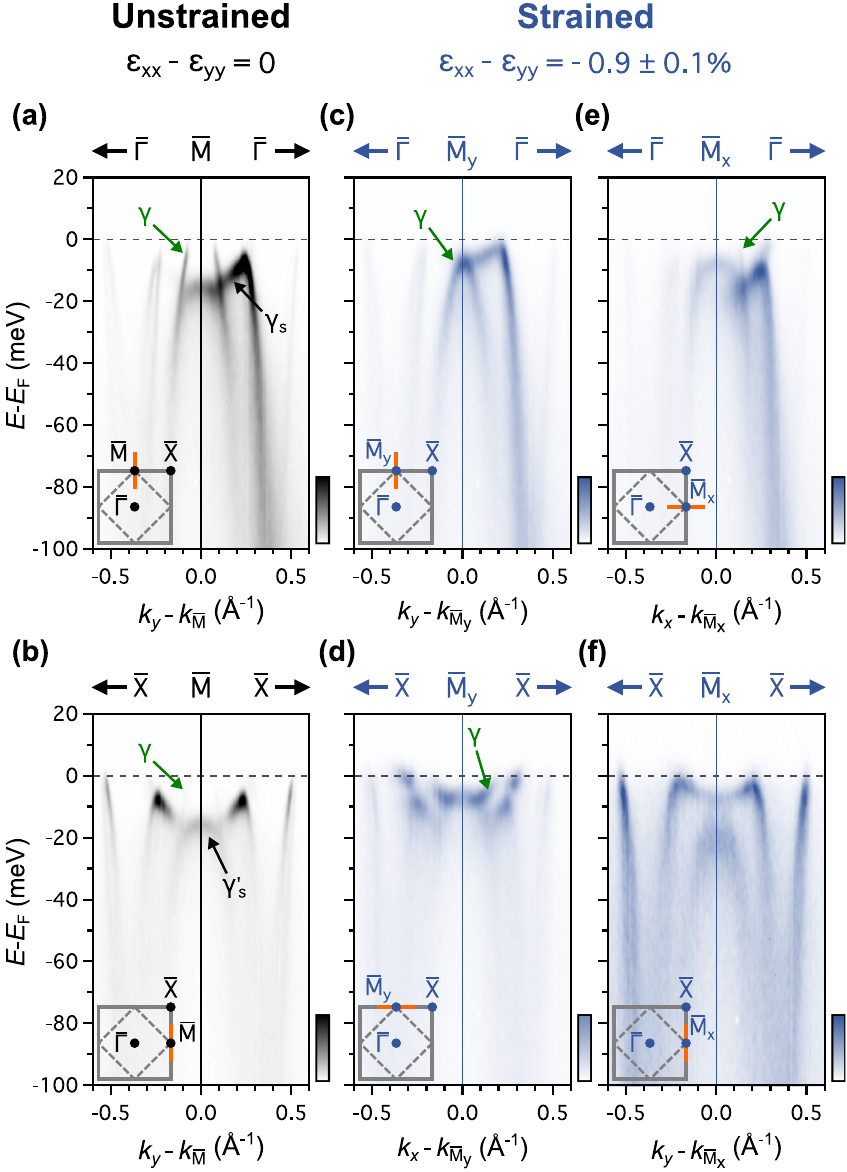}
\caption{\label{fig:fig2} Dispersions ($h\nu$=40eV, LH-pol.) close to the $\overline{\mathrm{M}}$-point of unstrained Sr\textsubscript{2}RuO\textsubscript{4} measured along the (a) $\overline{\mathrm{\Gamma}}$-$\overline{\mathrm{M}}$ and (b) $\overline{\mathrm{M}}$-$\overline{\mathrm{X}}$ directions. (c-f) Equivalent dispersions measured along the (c) $\overline{\mathrm{\Gamma}}$-$\overline{\mathrm{M}}_y$ (d) $\overline{\mathrm{M}}_y$-$\overline{\mathrm{X}}$, (e) $\overline{\mathrm{\Gamma}}$-$\overline{\mathrm{M}}_x$ and (f) $\overline{\mathrm{M}}_x$-$\overline{\mathrm{X}}$ direction for a  strained sample (${\varepsilon_{xx}-\varepsilon_{yy}=-0.9\pm0.1\%}$).}
\end{figure}

We show in Fig.~\ref{fig:fig2}(a,b) the band dispersions of unstrained Sr\textsubscript{2}RuO\textsubscript{4}, measured along the high-symmetry $\overline{\mathrm{\Gamma}}$-$\overline{\mathrm{M}}$ and $\overline{\mathrm{M}}$-$\overline{\mathrm{X}}$ directions. While distinct directions in the bulk, these are formally equivalent paths in the surface Brillouin zone (see insets). Nonetheless, the ARPES matrix elements vary significantly for measurements performed along these directions, and we will thus refer throughout to the conventional symmetry points of the surface Brillouin zone, with $\overline{\mathrm{M}}$ located at the $(\pi/a,0)$ or $(0,\pi/a)$ points of the tetragonal Brillouin zone, and $\overline{\mathrm{X}}$ at $(\pi/a,\pi/a)$. Along $\overline{\mathrm{\Gamma}}$-$\overline{\mathrm{M}}$, the hole band crossing $E_\mathrm{F}$ closest to the $\overline{\mathrm{M}}$-point in Fig.~\ref{fig:fig2}(a) is the bulk $\gamma$ band (Fig.~\ref{fig:fig1}(b)), which is predominantly derived from $d_{xy}$ orbitals. For such a two-dimensional $d_{xy}$ band, a saddle point is expected at the $\overline{\mathrm{M}}$ point of the Brillouin zone (Fig.~\ref{fig:fig1}(c,d)). While first-principles calculations suggest that its associated van Hove singularity (vHS) should be located more than 60 meV above the Fermi level \cite{Autieri2014}, electronic correlations renormalize this to only $\approx14$~meV above $E_\mathrm{F}$ \cite{Shen2007a,Tamai2019a,Burganov2016a}. Consistent with previous measurements \cite{Veenstra2013}, we find that a very weak replica of this band is also visible backfolded to the $\overline{\mathrm{M}}$-$\overline{\mathrm{X}}$ direction (Fig.~\ref{fig:fig2}(b)) due to the surface octahedral rotations. 

Additional surface states are also evident. The saddle point of the surface $\gamma$-band (SP$_1$)
 is pushed below the Fermi level~\cite{Shen2001} in the lower screening environment of the surface, with small additional downward shifts from band narrowing due to the octahedral rotation of the surface layer (see Supplementary Fig.~S5~\cite{SM}). Moreover, the $\overline{\mathrm{\Gamma}}$-$\overline{\mathrm{M}}$ and $\overline{\mathrm{M}}$-$\overline{\mathrm{X}}$ directions are folded onto each other by the doubling of the surface unit cell (Fig.~\ref{fig:fig1}(g,h)). Experimentally, the signatures of this are visible in our measured dispersions in Figs.~\ref{fig:fig2}(a,b) as a degeneracy at $\overline{\mathrm{M}}$ of the electron- ($\gamma_s$) and hole-like ($\gamma_s'$) surface $\gamma$ bands, located at a binding energy of 16~meV. The latter branch is most strongly visible along the $\overline{\mathrm{M}}$-$\overline{\mathrm{X}}$ direction (Fig.~\ref{fig:fig2}(b)), while the upward dispersing branch is clearly seen in the $\overline{\mathrm{\Gamma}}$-$\overline{\mathrm{M}}$ measurements (Fig.~\ref{fig:fig2}(a)). 

Interestingly, where $\gamma_s$ crosses the surface $\beta$ band ($\beta_s$), our tight-binding modelling (Fig.~\ref{fig:fig1}(h)) indicates that a small hybridisation gap is opened by spin-orbit coupling (SOC, inset of Fig.~\ref{fig:fig1}(h), see also Supplemental Fig.~S3~\cite{SM}). The resulting band hybridisation causes the formation of a new saddle point for the upper branch (SP$_2$ in Fig.~\ref{fig:fig1}(g,h)) while the lower branch develops a local band maximum. In our measurements of the surface electronic structure shown in Fig.~\ref{fig:fig2}(a,b), only the lower branch is visible in the occupied states, forming M-shaped bands along both $\overline{\mathrm{M}}$-$\overline{\mathrm{\Gamma}}$ and $\overline{\mathrm{M}}$-$\overline{\mathrm{X}}$, which are gapped from the Fermi level by $7\pm{2}$~meV. 

Significant changes in the electronic structure occur with uniaxial compression along the bulk Ru-O ($x$) direction (see also Supplementary Fig.~S6~\cite{SM}). $k_F$ of the bulk $\gamma$ band is increased along the direction of applied compressive strain (we denote this as $\overline{\mathrm{\Gamma}}$-$\overline{\mathrm{M}}_x$, Fig.~\ref{fig:fig2}(e)), while the $\gamma$ band is pushed down below the Fermi level along the perpendicular $\overline{\mathrm{\Gamma}}$-$\overline{\mathrm{M}}_y$ direction (Fig.~\ref{fig:fig2}(c)). The band top along $\overline{\mathrm{\Gamma}}$-$\overline{\mathrm{M}}_y$, and thus the position of its associated vHS, is now located $8\pm2$~meV below $E_\mathrm{F}$, confirming our previous observation of a strain-induced bulk Lifshitz transition in Sr$_2$RuO$_4$~\cite{Sunko2019b}.

The evolution of the surface electronic structure is more complex. Along the $\overline{\mathrm{\Gamma}}$-$\overline{\mathrm{M}}_y$ direction (Fig.~\ref{fig:fig2}(c), also visible along the symmetry-equivalent $\overline{\mathrm{X}}$-$\overline{\mathrm{M}}_x$ direction, Fig.~\ref{fig:fig2}(f)), the M-shaped band of the unstrained surface electronic structure (Fig.~\ref{fig:fig2}(a,b)) is pushed upwards, reaching almost to the Fermi level. In contrast, along $\overline{\mathrm{\Gamma}}$-$\overline{\mathrm{M}}_x$ (Fig.~\ref{fig:fig2}(e), and most clearly seen along the symmetry-equivalent $\overline{\mathrm{X}}$-$\overline{\mathrm{M}}_y$ direction, Fig.~\ref{fig:fig2}(d)), the same M-shaped band is pushed down, breaking the $C_4$ symmetry of the unstrained surface and leading to the initially unoccupied branch (Fig.~\ref{fig:fig1}(h)) moving below the Fermi level. A spin-orbit hybridisation gap of $\approx4$~meV is now visible between the surface $\gamma_s$ and $\beta_s$ bands, centered $\approx5$~meV below the Fermi level. 

\begin{figure*}
\includegraphics{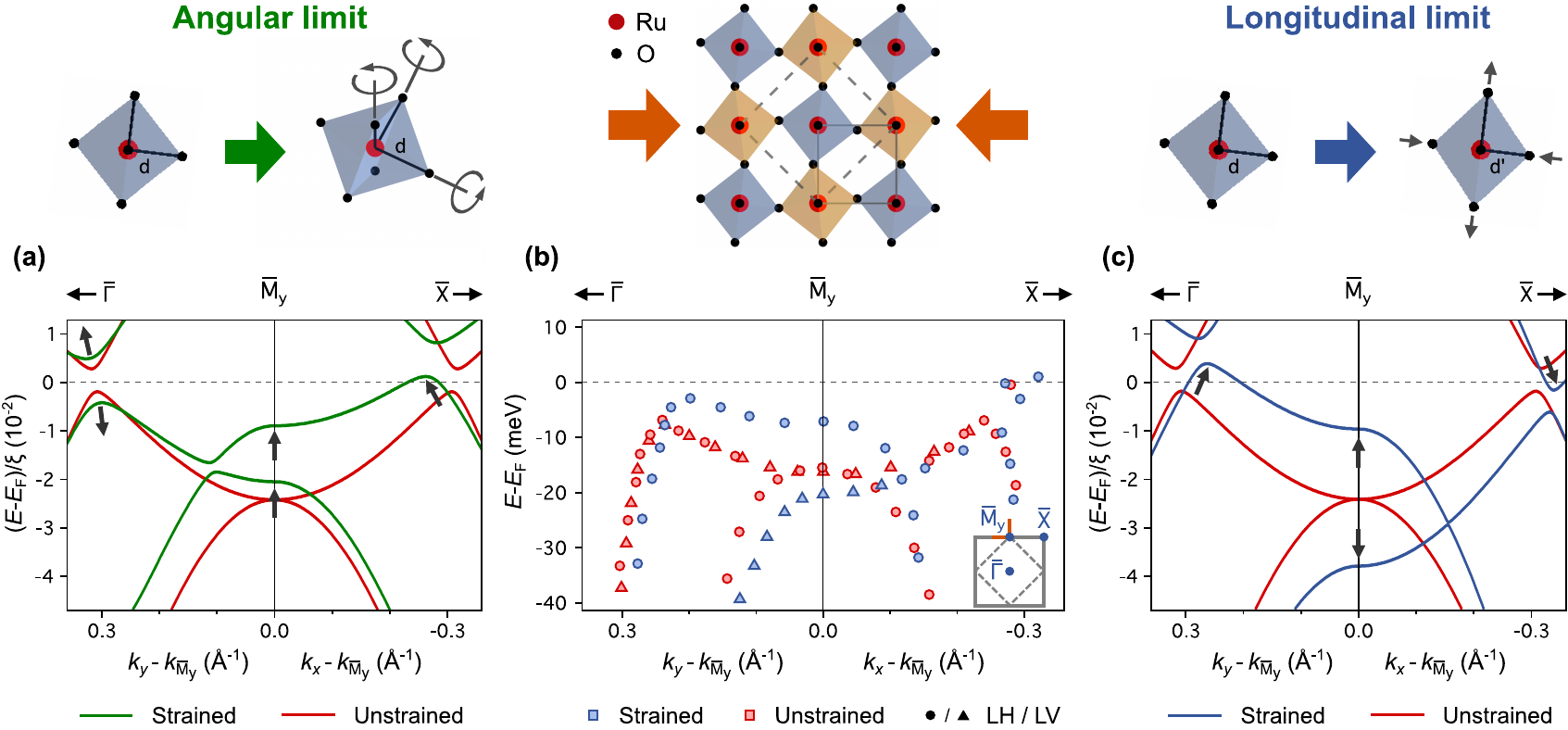}
\caption{\label{fig:fig3} Evolution of the surface electronic structure with uniaxial compression in the (a) angular and (c) longitudinal limits (see text). The dispersions extracted from our measured ARPES data are shown in (b). The calculations employ a $B_{1g}$ strain of ${\varepsilon_{xx}-\varepsilon_{yy}=-2.4\%}$, overestimating the experimental value as is also the case for bulk calculations~\cite{Barber2019,acharya_evening_2019} (see Supplementary Material~\cite{SM}).}
\end{figure*}

To help visualise these strain-dependent changes, we show in Fig.~\ref{fig:fig3} the surface band dispersions along the $\overline{\mathrm{\Gamma}}$-$\overline{\mathrm{M}}_y$-$\overline{\mathrm{X}}$ direction. The dispersions in Fig.~\ref{fig:fig3}(b) are extracted from measurements performed using both linear horizontal (Fig.~\ref{fig:fig2}(c,d)) and vertical (Supplemental Fig.~S7~\cite{SM}) light polarisation, where modified transition matrix elements better highlight the different band features (see also Supplementary Fig.~S8~\cite{SM} for equivalent surface band dispersions extracted along the symmetry-equivalent $\overline{\mathrm{X}}$-$\overline{\mathrm{M}}_x$-$\overline{\mathrm{\Gamma}}$ direction where the different experimental geometry again leads to distinct matrix elements). As well as confirming the surface Lifshitz transitions discussed above, these highlight an additional splitting of the originally 4-fold degenerate vHS derived from the backfolded bands at $\overline{\mathrm{M}}$ into two distinct 2-fold degenerate saddle points, with the two branches split by $\approx12$~meV.

Our extracted dispersions thus point to a strong breaking of $C_4$ symmetry at the surface. This is naturally expected given the anisotropic strain; the details of how this reshapes the electronic structure, however, are less obvious. In the bulk, the effect of uniaxial stress is well understood in terms of a simple compression of the RuO$_6$ octahedra in the direction of the applied stress, with a corresponding bond-length expansion in the perpendicular direction due to the Poisson effect. At the surface, however, the RuO$_6$ octahedra are already rotated around the $c$-axis in the absence of strain. The most natural starting assumption would therefore be that strain is accommodated by further rotations and tilts of these octahedra --- we term this the {\it angular limit}. Assuming perfectly rigid octahedra, the rotations required to accommodate the strain are uniquely defined, and require a combination of in-plane rotation and out-of-plane octahedral tilting (see Supplemental Material~\cite{SM} and Figs.~S9 and S10). From the resulting fully-constrained changes in the geometrical configuration, we can directly calculate modifications of the inter-orbital hoppings within our tight-binding model,  allowing us to predict the influence of the strain accommodation on the surface electronic structure without the introduction of any additional free parameters. We show the results of this in Fig.~\ref{fig:fig3}(a). 

While the lowering of the symmetry of the surface electronic structure from $C_4$ to $C_2$ is, of course, reproduced by this model, we find that the strain-mediated changes in the electronic structure are otherwise in qualitative disagreement with our experimental measurements (Fig.~\ref{fig:fig3}(b)). The top of the occupied M-shaped band is pushed upwards towards the Fermi level along $\overline{\mathrm{M}}_y$-$\overline{\mathrm{X}}$, rather than the downwards shift that is required to reproduce the surface Lifshitz transition observed experimentally. Meanwhile, along $\overline{\mathrm{\Gamma}}$-$\overline{\mathrm{M}}_y$, the surface bands develop a strong hybridisation gap, pushing the occupied states down well below the Fermi level, again in contrast to our experimental observations (Fig.~\ref{fig:fig3}(b)). Finally, while the 4-fold degenerate vHS at $\overline{\mathrm{M}}$ does become split under strain, both branches are split off above its position for the unstrained surface, distinct to the experimental situation where the new saddle points are split almost symmetrically about the unstrained case.

On the other hand, if we consider a \textit{longitudinal limit}, where the surface octahedra are only able to distort via bond-length deformations, we predict an electronic structure which is in excellent agreement with our measured dispersions (Fig.~\ref{fig:fig3}(c)). We thus conclude that application of uniaxial pressure to the bulk crystal leads, at least dominantly, to a change in Ru-O bond length of the surface octahedra. 

\begin{figure}[t]
\includegraphics{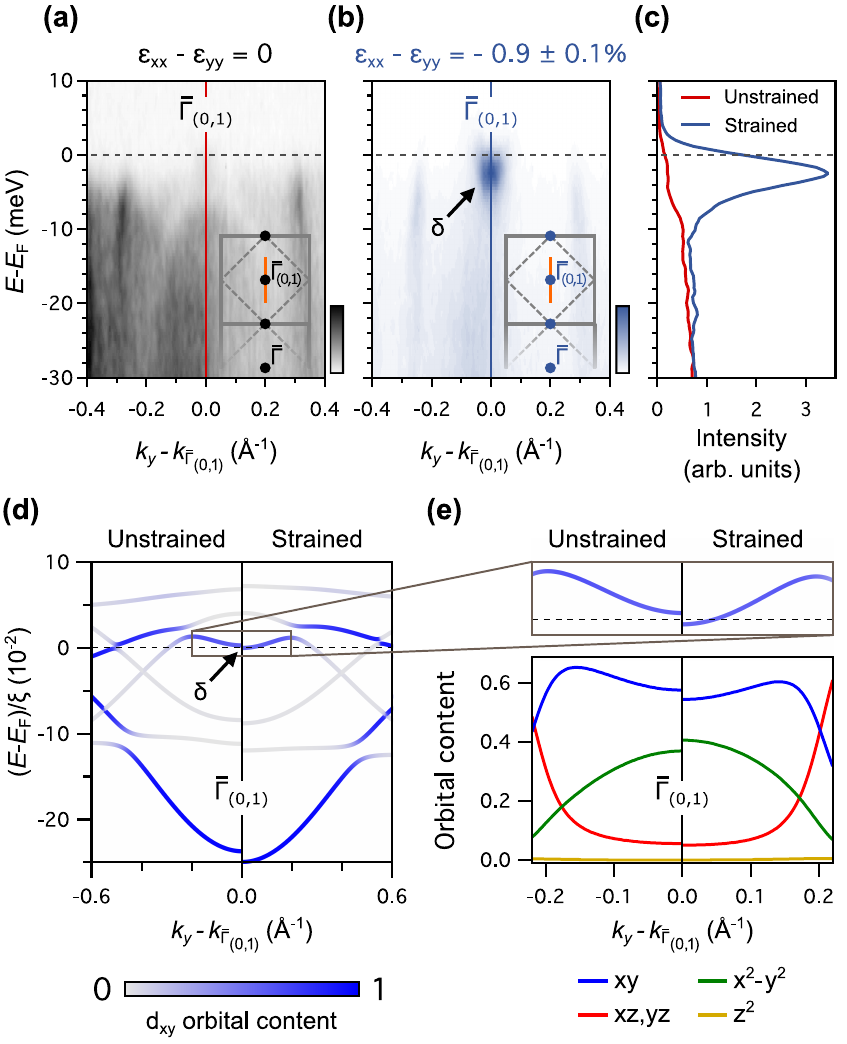}
\caption{\label{fig:fig4} Measured dispersions ($h\nu$=40eV, LV-pol.) centered at the $\overline{\mathrm{\Gamma}}$-point of the second Brillouin zone ($\overline{\mathrm{\Gamma}}_{(0,1)}$) for (a) unstrained and (b) strained Sr\textsubscript{2}RuO\textsubscript{4}, and (c) $\overline{\mathrm{\Gamma}}$-point EDCs. (d) Tight-binding calculations showing the effect of strain within the longitudinal limit on the states near the Brillouin zone center, with projected $d_{xy}$ orbital weight. (e) Magnified view of band dispersions in the vicinity of the new $\delta$ pocket (top), and corresponding $d$-orbital content (bottom).}
\end{figure}

We show in Fig.~\ref{fig:fig4} how such strain-driven bond-length distortions additionally create a new Fermi pocket at the Brillouin zone centre. We label this $\delta$, in analogy with the corresponding $\Gamma$-centered Fermi pocket in Sr$_3$Ru$_2$O$_7$ \cite{Tamai2008a}. Our tight-binding modelling (Fig.~\ref{fig:fig4}(d,e)) indicates that this $\delta$ band has predominantly $d_{xy}$ and $d_{x^2-y^2}$ orbital character. The $d_{x^2-y^2}$ band is part of the $e_g$ manifold, split off above the $t_{2g}$ states by a large octahedral crystal field. For bulk Sr$_2$RuO$_4$, its hybridisation with $d_{xy}$ orbitals in the $t_{2g}$ manifold is forbidden by symmetry. In the surface layer, however, the octahedral rotation permits their mixing (see Supplementary Fig.~S5~\cite{SM}), leading to a local depression at the top of the backfolded surface $\gamma$ band at $\overline{\mathrm{\Gamma}}$. Consistent with prior work \cite{Shen2001}, our measurements of the unstrained sample indicate that the bottom of the resulting $\delta$ pocket is above the Fermi level. Our calculations, however, show that the $d_{xy}$/$d_{x^2-y^2}$ orbital mixing is enhanced under strain (Fig.~\ref{fig:fig4}(e)), lowering the energy of the bottom of the $\delta$ band (Fig.~\ref{fig:fig4}(d,e)), and in turn driving another Lifshitz transition leading to the creation of a new $\delta$-pocket Fermi surface as observed experimentally (Fig.~\ref{fig:fig4}(a-c)).

The fact that bond-length changes appear to dominate the structural response to an applied uniaxial stress here may, at first sight, appear surprising, given the pre-existing surface reconstruction and the propensity of perovskite-type oxides to structural distortions involving octahedral rotations~\cite{Koster2012,Markovic2020,Torrance1992}. We note, however, that a Lifshitz transition itself can be expected to give a contribution to the electronic component of the compressibility \cite{Lifshitz1960}, softening the lattice in line with the required bond-length changes that we find to dominate the structural distortions here. The hierarchy of Lifshitz transitions observed here under strain thus potentially provides an electronic incentive to favour bond-length distortion over rigid octahedral rotation, and motivates future study of the detailed strain-dependent distortions from surface-sensitive structural probes and first-principles calculations of surface structure under strain. Furthermore, we note that many of the other Ruddlesden-Popper ruthenates (and many perovskites in general) host octahedral rotations in their bulk crystal structure. Our findings thus motivate future studies for how strain -- which can have a striking influence on their collective states~\cite{Brodsky17,Ricco2018a,McLeod19,Dashwood22} --  modifies not just lattice constants, but also the local crystal structure in these systems. Beyond bulk systems, this is of interest for the study of epitaxial thin films, where biaxial strain can readily be coupled from a growth substrate, offering further opportunities for control~\cite{Burganov2016a}.

Already at the surface, it may be possible to realise some of the rich phenomenology of the bulk systems using strain as a tuning parameter. In Sr$_3$Ru$_2$O$_7$, for example, field-tuning of near-$E_\mathrm{F}$ vHSs, similar to those studied here, to the Fermi level is thought to drive the emergence of quantum criticality~\cite{Grigera2001,Borzi2007} and the stabilization of spin-density-wave phases~\cite{Lester2015}. Recent scanning tunneling microscopy measurements suggest that magnetic fields as high as 32~T would be required to achieve similar field-tuned Lifshitz transitions for the surface layer of Sr$_2$RuO$_4$~\cite{Marques2021a}, while we have found here that the corresponding Lifshitz transition is naturally driven by modest applied uniaxial pressure. Moreover, we find that the M-shaped surface band which is pushed towards the Fermi level becomes flatter under the resulting strain (Fig.~\ref{fig:fig2}(c)), potentially mediating a crossover to a so-called higher (fourth) order singularity, characterised by a power-law divergence in its associated density of states~\cite{Efremov2019}. Such a `multicritical' singularity has been proposed as key to explaining the exotic collective states of the sister compound Sr$_3$Ru$_2$O$_7$. Our study, whereby a hierarchy of surface Lifshitz transitions are induced and tuned by an applied uniaxial stress, raises the tantalising prospect that the surface of Sr$_2$RuO$_4$ could be driven to host its own quantum critical states, providing new possibilities for studying such phases with spectroscopic approaches.  

{\it Acknowledgements:} We thank J.~Betouras, A.~Chandrasekaran, D.~Halliday, C.~Marques, L.~Rhodes, A.~Rost, V.~Sunko, and P.~Wahl for useful discussions. We gratefully acknowledge support from the Engineering and Physical Sciences Research Council (Grant Nos.~EP/T02108X/1 and EP/R031924/1), the European Research Council (through the QUESTDO project, 714193), and the Leverhulme Trust (Grant No.~RL-2016-006). E.A.M., A.Z., and I.M. gratefully acknowledge studentship support from the International Max-Planck Research School for Chemistry and Physics of Quantum Materials. N.K. is supported by a KAKENHI Grants-in-Aids for Scientific Research (Grant Nos. 18K04715, and 21H01033), and Core-to-Core Program (No. JPJSCCA20170002) from the Japan Society for the Promotion of Science (JSPS) and by a JST-Mirai Program (Grant No. JPMJMI18A3). APM and CWH acknowledge support from the Deutsche Forschungsgemeinschaft - TRR 288 - 422213477 (project A10). We thank Diamond Light Source for access to Beamline I05 (Proposals SI27471 and SI28412), which contributed to the results presented here. The research data supporting this publication can be accessed at https://doi.org/10.17630/be3be544-f107-4863-93a3-eff656095c15~\cite{data_ref}.

%


\definecolor{orange}{RGB}{255,127,0}
\definecolor{blue2}{RGB}{33,114,173}
\renewcommand\thefigure{S\arabic{figure}} 
\let\oldAA\AA
\renewcommand{\AA}{\text{\normalfont\oldAA}}

\setcounter{figure}{0}

\newpage

\onecolumngrid

\section*{Supplemental material}


\section{Strain characterization}

\

\begin{figure}[th]
\includegraphics{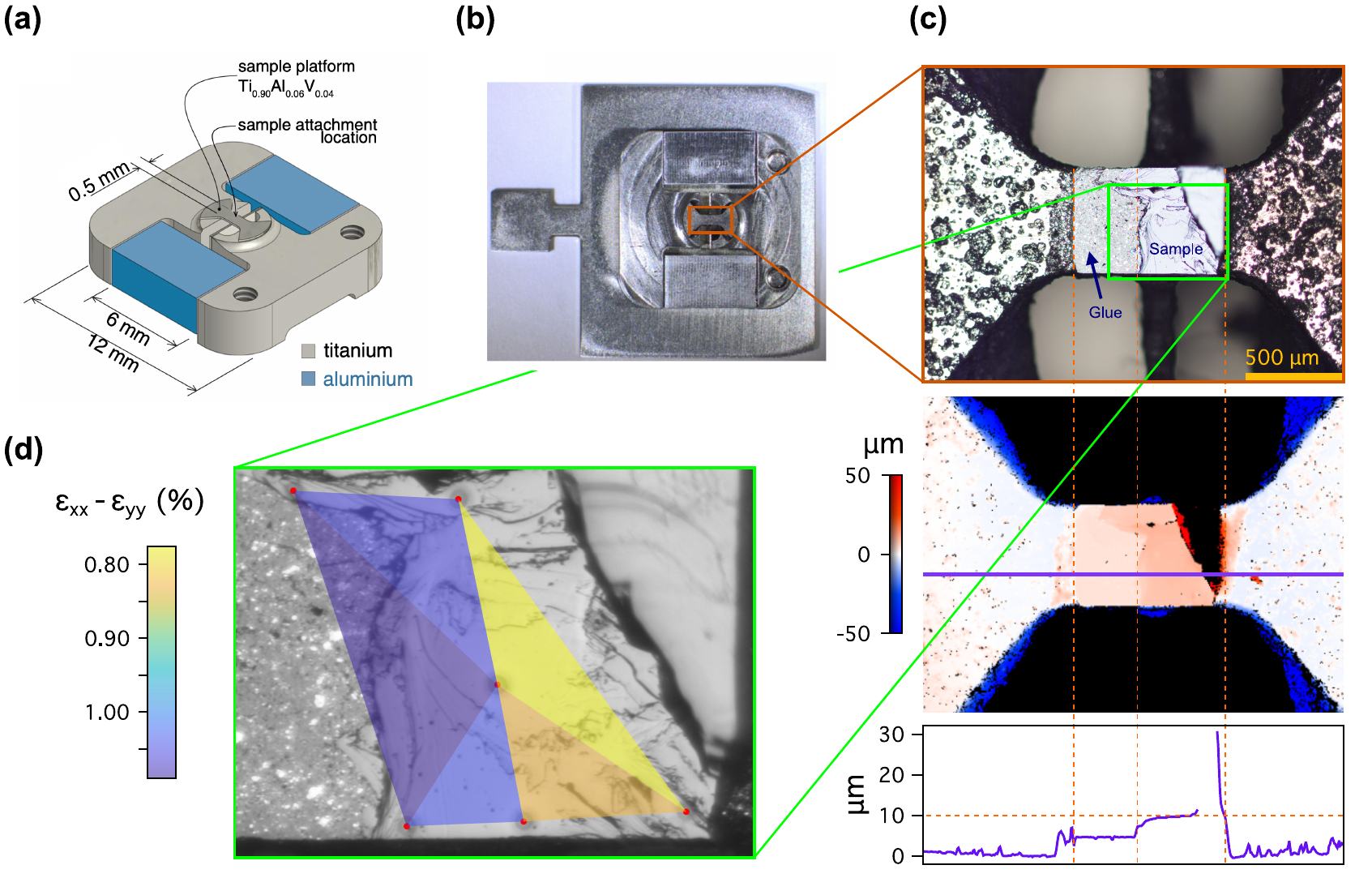}
\caption{\label{fig:figS1} (a) Illustration of the differential thermal contraction strain-rig. (b) Photograph of the strain-rig mounted on a standard ARPES flag plate. (a,b) are reproduced from Ref.~\cite{Sunko2019b}. (c) The Sr$_2$RuO$_4$ sample for the strain-ARPES measurements shown in the main text, following sample cleaving. Profilometry analysis shows that the measured region of the sample has a thickness of $\approx\!5\,\mu{\rm m}$ above the silver-epoxy layer. (d) Optical post-characterisation of the strain. Focused ion beam milling was used to create a pattern of dots on the sample surface, which were imaged with optical microscopy as the sample was cooled in a flow cryostat. Finite element analysis shows a $B_{1g}$ strain of ${\varepsilon_{xx}-\varepsilon_{yy}=-0.9\pm0.1\%}$ was achieved in the sample, with a slight increase towards the left hand edge of the sample. }
\end{figure}

\newpage

\section{Tight-binding model}

\begin{figure}[!b]
\includegraphics{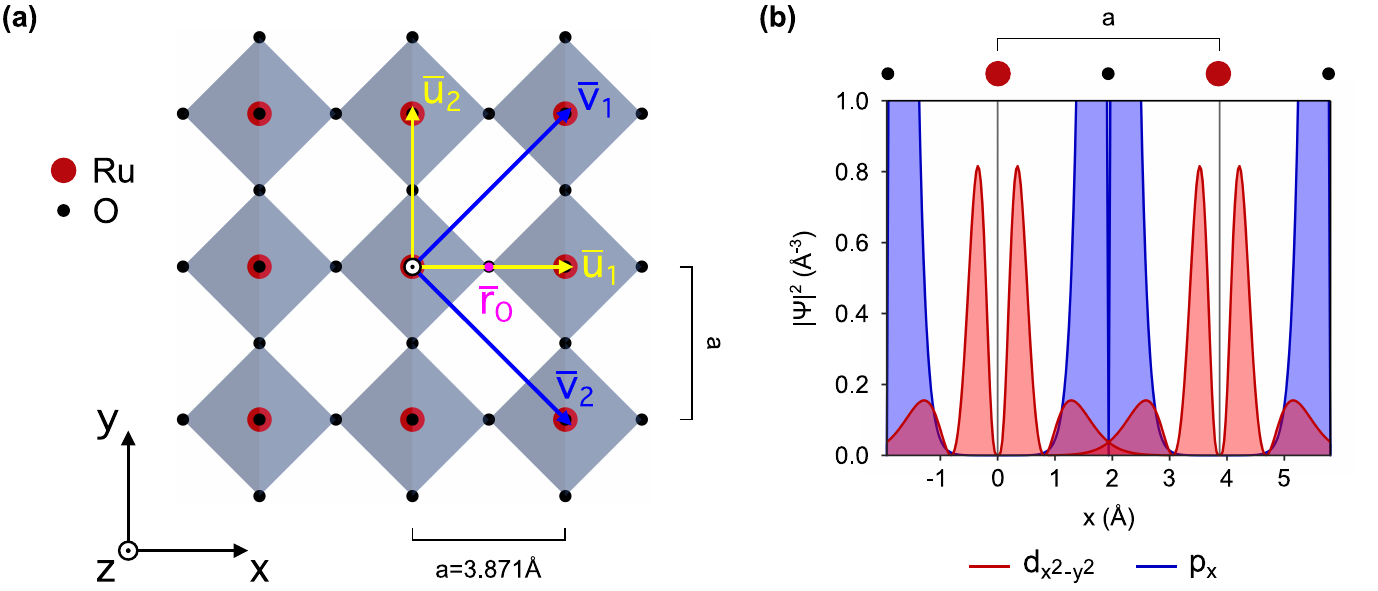}
\caption{\label{fig:figS2} (a) Top view of a bulk RuO\textsubscript{2} layer of Sr\textsubscript{2}RuO\textsubscript{4}. The nearest-neighbor $(\vec{u}_1,\vec{u}_2)$ and next-nearest-neighbor $(\vec{v}_1,\vec{v}_2)$ hopping vectors are indicated. (b) Example radial wavefunctions along the Ru-O-Ru bond direction, calculated for core charges of $Z_{Ru}=7.8$, $Z_{O}=5.9$, illustrating the calculation of the overlap integrals within our tight-binding model.}
\end{figure}

To model the electronic structure of Sr$_2$RuO$_4$, we adopt a tight-binding model. In our model we replace the actual tight-binding hopping integrals by overlaps (or products of overlaps) between normalized atomic orbitals centered on specific pairs of nearby sites. These overlaps are dimensionless, and so we cannot make predictions in energy units. Nonetheless, due to the central nature of the Coulomb potential, the symmetry properties of the overlaps and their response to crystal distortions mimic those of the true tight-binding hopping integrals and for our purposes they typically suffice. Explicitly, we use the $\{\ket{d_{xy}},\ket{d_{yz}},\ket{d_{z^2}},\ket{d_{xz}},\ket{d_{x^2-y^2}}\}$ $4d$-orbital basis at the Ru sites and the $\{\ket{p_{x}},\ket{p_{y}},\ket{p_{z}}\}$ $2p$-orbital basis at the O sites. Each basis element can be expressed as $\Psi_{nlm}=R_{nl}Y_{lm}$, where $R$ is the radial wavefunction, $Y$ is a real spherical harmonic, $(n,l)$ are standard hydrogenic quantum numbers, and $m$ labels the orbital flavor. At the Ru and O sites, the radial dependence is given by:
\begin{equation}
R_{42}(r)=\left(\frac{Z_{Ru}}{a_0}\right)^{3/2}e^{-\rho/2}\frac{1}{96\sqrt{5}}(6-\rho)\rho^2
\end{equation}
\begin{equation}
R_{21}(r)=\left(\frac{Z_O}{a_0}\right)^{3/2}e^{-\rho/2}\frac{1}{2\sqrt{6}}\rho,
\end{equation}
where $a_0$ is the Bohr radius and $\rho=\frac{2Zr}{a_0n}$. Thus, the spatial extent of the Ru and O orbitals is characterized by effective nuclear charges $Z_{\rm Ru}$ and $Z_{\rm O}$, which are the main phenomenological parameters in our model (see, e.g. Fig.~\ref{fig:figS2}(b)). The Hamiltonian includes in-plane nearest-neighbor (NN) Ru-O-Ru hopping connecting inequivalent Ru sites and direct next-nearest-neighbor (NNN) Ru-Ru hopping between equivalent Ru sites. In the former case the hopping parameters are obtained by numerical integration of:
\begin{equation}\label{eq1}
t_{NN}^{mm'}=\sum_{m_O}\bra{4d_m(\vec{r})}\ket{2p_{m_O}(\vec{r}+\vec{r}_O)}\bra{2p_{m_O}(\vec{r}+\vec{r}_O))}\ket{4d_{m'}(\vec{r}+\vec{u})},
\end{equation}
where $\vec{u}$ is a nearest-neighbor vector along the Ru atoms and $\vec{r}_O$ is the O site in between (Fig. \ref{fig:figS2}). Similarly, in the latter case they are obtained by computing:
\begin{equation}\label{eq2}
t_{NNN}^{mm'}=\bra{4d_m(\vec{r})}\ket{4d_{m'}(\vec{r}+\vec{v})},
\end{equation}
where $\vec{v}$ is a next-nearest-neighbor vector (Fig. \ref{fig:figS2}). A diagonal crystal field matrix ${\bf H}_{\rm CF}$ was added to the Hamiltonian to separate the $t_{2g}$ and $e_{g}$ manifolds of the $4d$-orbital basis and to shift the energy of the $d_{xy}$ orbital within the $t_{2g}$ manifold, as expected for the tetragonal structure. A similar splitting of the orbitals in the $e_g$ manifold would be expected, but we neglect these here for simplicity as we have no unique way to set its value in reference to the experiments. Additional octahedral rotations were included for the surface layer, in which case the orbitals at the Ru sites were rotated using the Wigner D-matrix formalism. On-site spin-orbit coupling (SOC) was included at the Ru sites, with the angular momentum operators also rotated to be consistent with any rotated octahedral configurations. 

The bulk Hamiltonian, where the absence of $c$-axis rotation makes the Ru sites equivalent (Fig. \ref{fig:figS2}), is given by:
\begin{equation}
{\bf H}_{\rm bulk} = \left( \begin{array}{cc}
1 & 0 \\ 0 & 1 \end{array} \right) \otimes \left( {\bf H}_{\rm CF} + {\bf H}_{\rm NN} + {\bf H}_{\rm NNN} \right) + {\bf H}_{\rm SOC},
\end{equation}
where the hopping matrices take the form:
\begin{equation}
{\bf H}_{\rm NN}=2\left[T_{{\vec u}_1}\cos(\vec{k}\cdot{\vec u}_1)+T_{{\vec u}_2}\cos(\vec{k}\cdot{\vec u}_2)\right]
\end{equation}
\begin{equation}
{\bf H}_{\rm NNN}=2\left[T_{{\vec v}_1}\cos(\vec{k}\cdot{\vec v}_1)+T_{{\vec v}_2}\cos(\vec{k}\cdot{\vec v}_2)\right],
\end{equation}
with the $T_{\vec{n}}$'s being $5 \times 5$ matrices whose entries are the overlap integrals calculated from equations (3) and (4), respectively. The crystal field matrix has the explicit form:
\begin{equation}
{\bf H}_{\rm CF} =
\begin{pmatrix}
c_{d_{xy}} & 0 & 0 & 0 & 0 \\
0     & 0 & 0 & 0 & 0 \\
0     & 0 & 0 & 0 & 0 \\
0     & 0 & 0 & c_{e_g} & 0 \\
0     & 0 & 0 & 0 & c_{e_g} \\
\end{pmatrix},
\end{equation}
where $c_{d_{xy}}$ shifts the energy of the $d_{xy}$ states within the $t_{2g}$ manifold and $c_{e_g}$ separates the $e_g$ states from the $t_{2g}$ states. The spin-orbit coupling is expressed as:
\begin{equation}
{\bf H}_{\rm SOC} = g\left[S^z\otimes{L^z}+\frac{1}{2}\left(S^+\otimes{L^-}+S^-\otimes{L^+}\right)\right],
\end{equation}
where $g$ is the coupling strength and the spin-1/2 and $l=2$ angular momentum operators are defined elsewhere \cite{Sakurai1994}. 

\begin{figure}[!t]
\includegraphics[width=0.8\columnwidth]{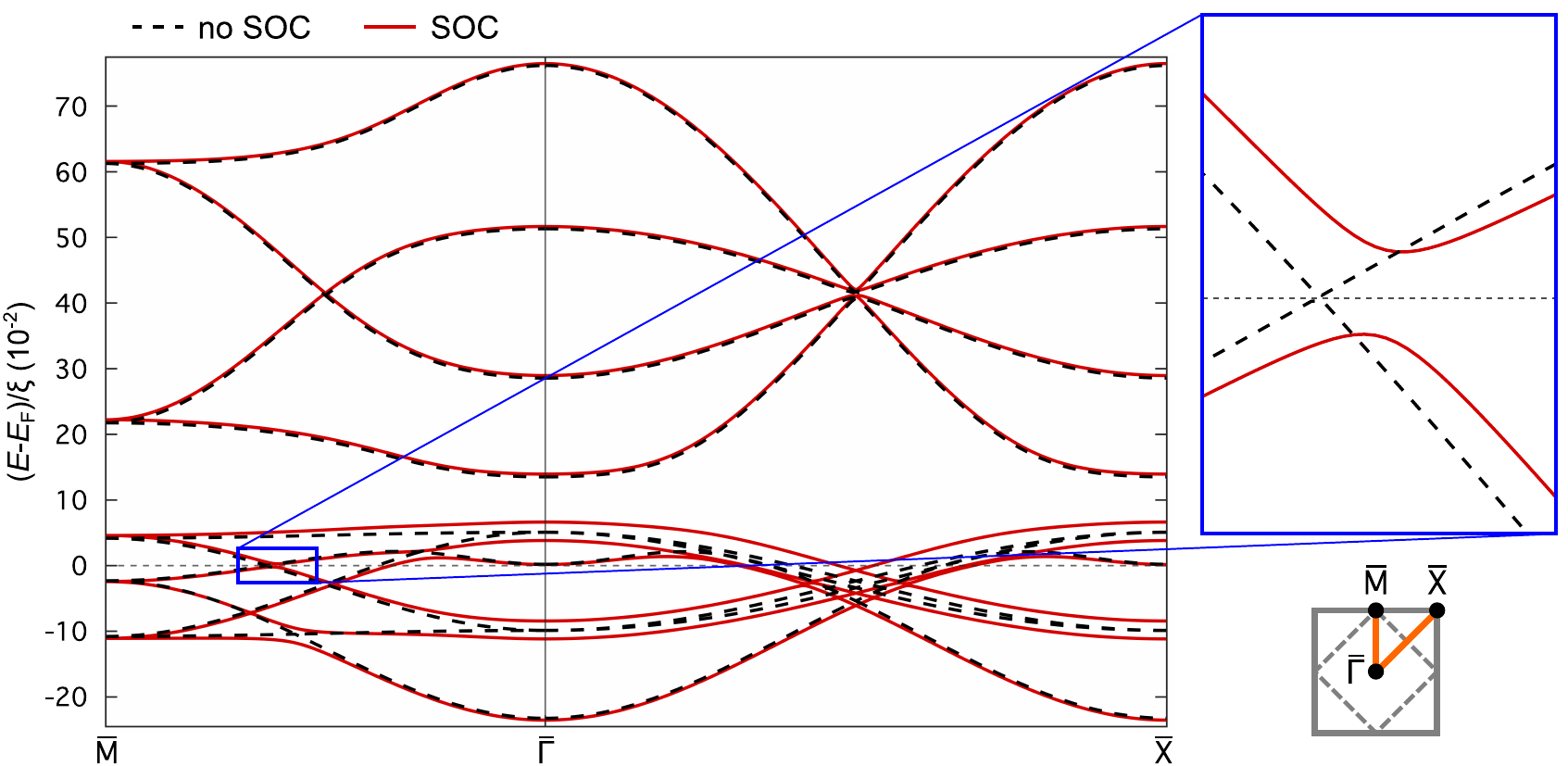}
\caption{\label{fig:figS3} The unstrained surface electronic structure of Sr$_2$RuO$_4$ along the $\overline{\mathrm{M}}$-$\overline{\mathrm{\Gamma}}$-$\overline{\mathrm{X}}$ direction calculated from our tight-binding model for a surface reconstruction consisting of a 7-degree rotation of the octahedra around the $c$-axis. The inset shows that spin-orbit coupling (SOC) drives the formation of a hybridization gap between the $\gamma_s$ and $\beta_s$ bands away from the $\overline{\mathrm{M}}$-point in the vicinity of $E_F$.}
\end{figure}

\begin{figure}[!t]
\includegraphics{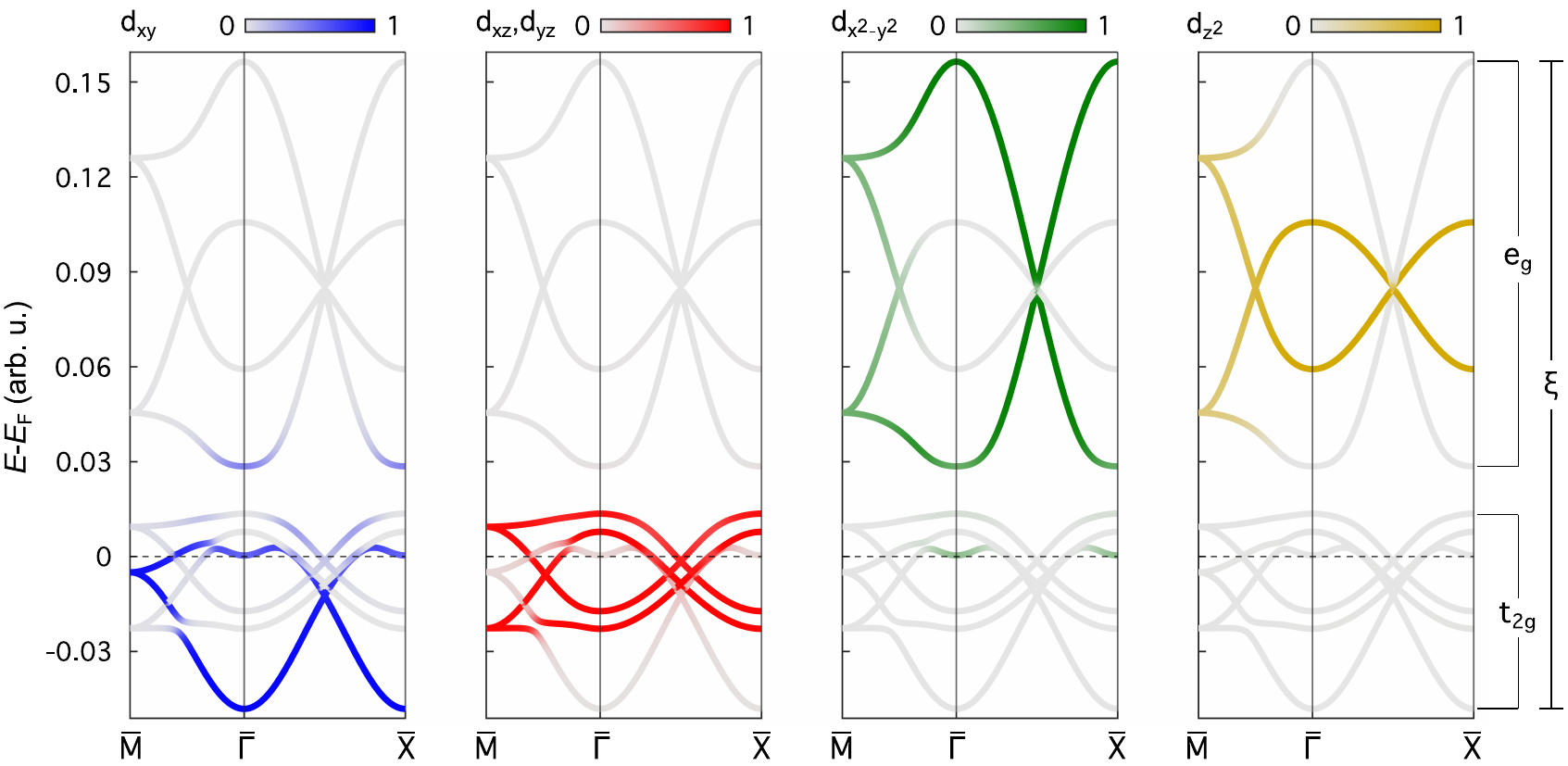}
\caption{\label{fig:figS4} Projected orbital content for the unstrained surface electronic structure of Sr$_2$RuO$_4$ along the $\overline{\mathrm{M}}$-$\overline{\mathrm{\Gamma}}$-$\overline{\mathrm{X}}$ direction calculated from our tight-binding model for a surface reconstruction consisting of a 7-degree rotation of the octahedra around the $c$-axis. The orbital content is projected to the rotated orbital basis, i.e., we work in the local orbital basis. The extent of the $t_{2g}$ and $e_g$ manifolds is indicated, as well as the total $d$-orbital bandwidth which we use as our normalization parameter, $\xi$.}
\end{figure}

\begin{figure}[!t]
\includegraphics[width=0.9\columnwidth]{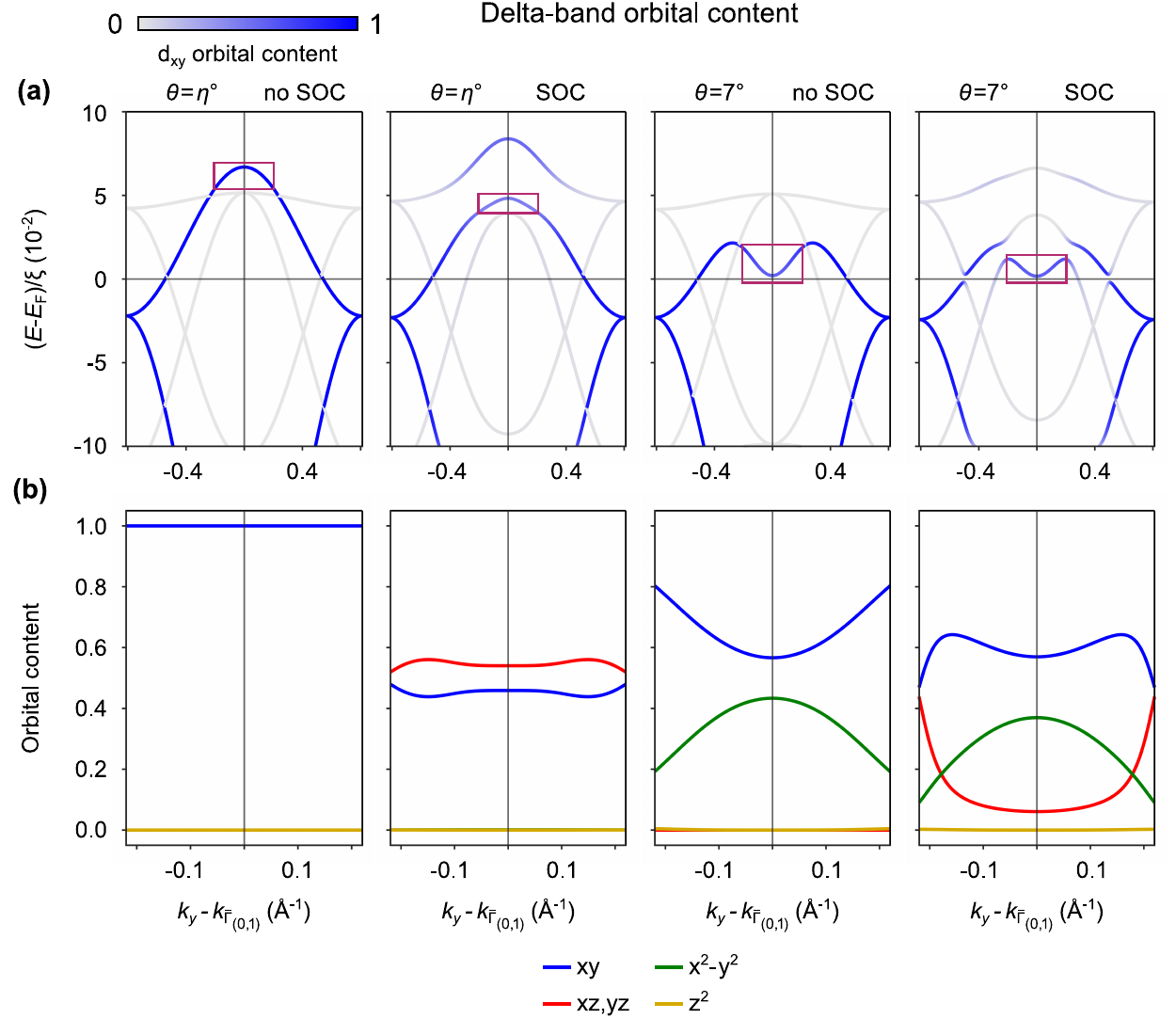}
\caption{\label{fig:figS5} (a) Calculated band structure for various combinations of octahedral rotations and spin-orbit coupling ($\eta$ indicates an infinitesimally small octahedral rotation angle). (b) Projected orbital content in the local orbital basis from the momentum region indicated in the purple boxes in (a), showing the influence of the rotations and SOC on the orbital mixing.}
\end{figure}

Similarly, the surface Hamiltonian is given by:
\begin{equation}
{\bf H}_{\rm surf} = \left( \begin{array}{cc}
1 & 0 \\ 0 & 1 \end{array} \right) \otimes \left[ \left( \begin{array}{cc}
1 & 0 \\ 0 & 1 \end{array} \right) \otimes {\bf H}_{\rm CF} + \left( \begin{array}{cc}
{\bf H}_{\rm NNN}^{\rm A} & {\bf H}_{\rm NN} \\ {\bf H}_{\rm NN}^{\rm \dagger} & {\bf H}_{\rm NNN}^{\rm B} \end{array} \right)\right] + {\bf H}_{\rm SOC},
\end{equation}
where the A (B) subscript refers to NNN processes between Ru A-sites (B-sites). The ${\bf H}_{\rm CF}$, ${\bf H}_{\rm NN}$ and ${\bf H}_{\rm NNN}$ matrices are defined as above. For the spin-orbit term the angular momentum operators must be rotated to match the octahedral rotations present at sites A and B: 
\begin{equation}
{\bf H}_{\rm SOC} = g\left[S^z \otimes
\begin{pmatrix}
L^z_A & 0 \\
0 & L^z_B \\
\end{pmatrix}
+\frac{1}{2}\left(S^+ \otimes
\begin{pmatrix}
L^-_A & 0 \\
0 & L^-_B \\
\end{pmatrix}
+S^- \otimes
\begin{pmatrix}
L^+_A & 0 \\
0 & L^+_B \\
\end{pmatrix}
\right)\right].
\end{equation}
We fix the Fermi level to the correct electron count of 4 electrons per Ru by numerically determining the Luttinger count. Because of the absence of a true energy scale in our model, we show the dispersions normalized by the total bandwidth of the surface band structure in the absence of uniaxial strain, which we denote by $\xi$. In Fig. \ref{fig:figS3} we show the surface electronic structure derived from our tight-binding model in the absence of uniaxial strain, including and excluding spin-orbit coupling. The hybridization gap between the $\gamma_s$ and $\beta_s$ surface bands occurs due to the spin-orbit mediated mixing of the $d_{xy}$ and $d_{xz/yz}$ orbitals from which the bands derive, respectively. In Fig. \ref{fig:figS4} we show the projected orbital content in the local orbital basis for the bands in Fig. \ref{fig:figS3}. We note a finite mixing of the $d_{x^2-y^2}$ states into the $t_{2g}$ manifold. To explore the origins of this further, we show in Fig.~\ref{fig:figS5} the calculated surface electronic structure with and without spin-orbit coupling and with octahedral rotations included vs. an infinitesimally small rotation, such that the bands are backfolded (i.e. we work with the 2-atom unit cell) but the orbitals remain unrotated. The local suppression of the top of the dominantly $d_{xy}$ band (highlighted by the purple box in the panels in Fig.~\ref{fig:figS5}(a)) is only found when octahedral rotations are included. Fig.~\ref{fig:figS5}(b) shows the projected orbital weight for this representative region. With no SOC and no rotations, the band here is of pure $d_{xy}$ character. Including spin-orbit coupling mixes in $d_{xz/yz}$ character, but is unable to induce any hybridization with the $d_{x^2-y^2}$ orbitals. In contrast, with octahedral rotations but no SOC, the $d_{xy}$ and $d_{x^2-y^2}$ orbitals hybridise strongly. When both rotations and SOC are included, the $d_{xy}$, $d_{xy/yz}$ and $d_{x^2-y^2}$ orbitals are all hybridized. Thus, while spin-orbit coupling leads to additional mixings between the $d_{xy}$ and $d_{xz/yz}$ orbitals, it is unable to hybridise the $t_{2g}$ and $e_g$ states: this, and the consequent mixing of $d_{xy}$ and $d_{x^2-y^2}$ character visible in Fig.~\ref{fig:figS4}, is generated by the octahedral rotation. 

\

The effects of strain are directly incorporated into our via the effects of the resulting geometrical distortions which in turn modify the overlap integrals defined in Equations~\ref{eq1} and~\ref{eq2}. We accommodate the strain either via bond-length changes (the \textit{longitudinal limit}) or via rotations of rigid octahedral (the \textit{angular limit}; see Rigid Octahedra Model below for a discussion of the relevant geometrical changes). For bulk Sr$_2$RuO$_4$, it is well established that single-particle calculations such as density-functional theory calculations significantly overestimate the required strain in order to drive the (bulk) Lifshitz transition, pointing to a significant role of electronic correlations in renormalising the Lifshitz strain to lower values~\cite{Barber2019,acharya_evening_2019}. Similarly, we find here that we require to include a $B_{1g}$ strain of ${\varepsilon_{xx}-\varepsilon_{yy}=-2.4\%}$ in our calculations in order to drive both the bulk and surface Lifshitz transitions, larger than the experimental strain by a similar factor to that found for previous bulk Sr$_2$RuO$_4$ calculations. We note that irrespective of the exact strain applied, the angular and longitudinal limits have qualitatively distinct behaviour in the strain-response of the surface electronic structure, and only the trends of the longitudinal limit are consistent with our experimental measurements.

\

\newpage

\

\section{Additional measurements}

\begin{figure}[!h]
\includegraphics[width=0.7\textwidth]{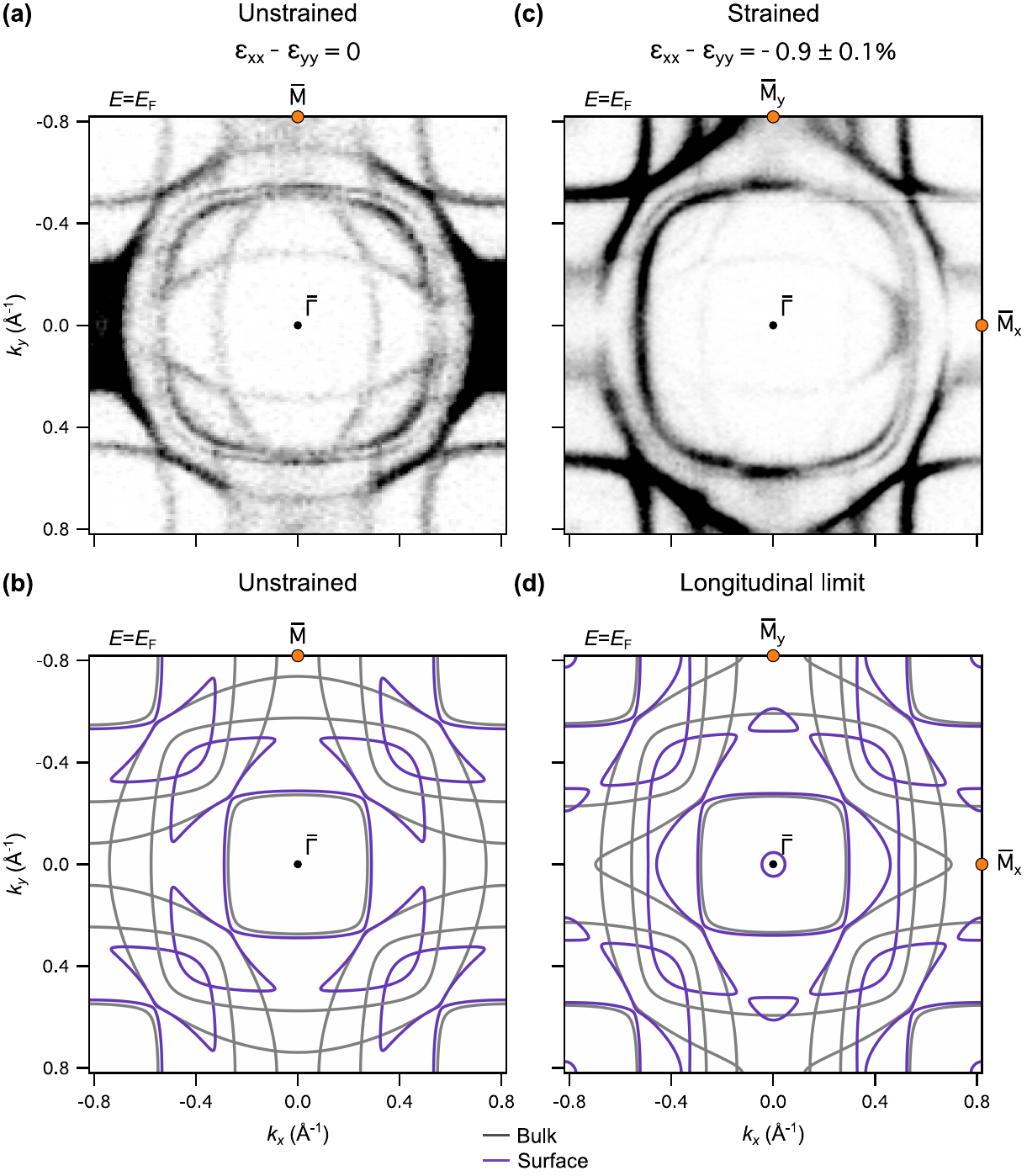}
\caption{\label{fig:figS6} (a) Measured ($h\nu=100$~eV) and (b) calculated Fermi surface for unstrained Sr$_2$RuO$_4$. (c,d) Equivalent measurements ($h\nu=68$~eV) and calculations for the strained case.}
\end{figure}

\

\begin{figure}[!h]
\includegraphics[width=0.86\textwidth]{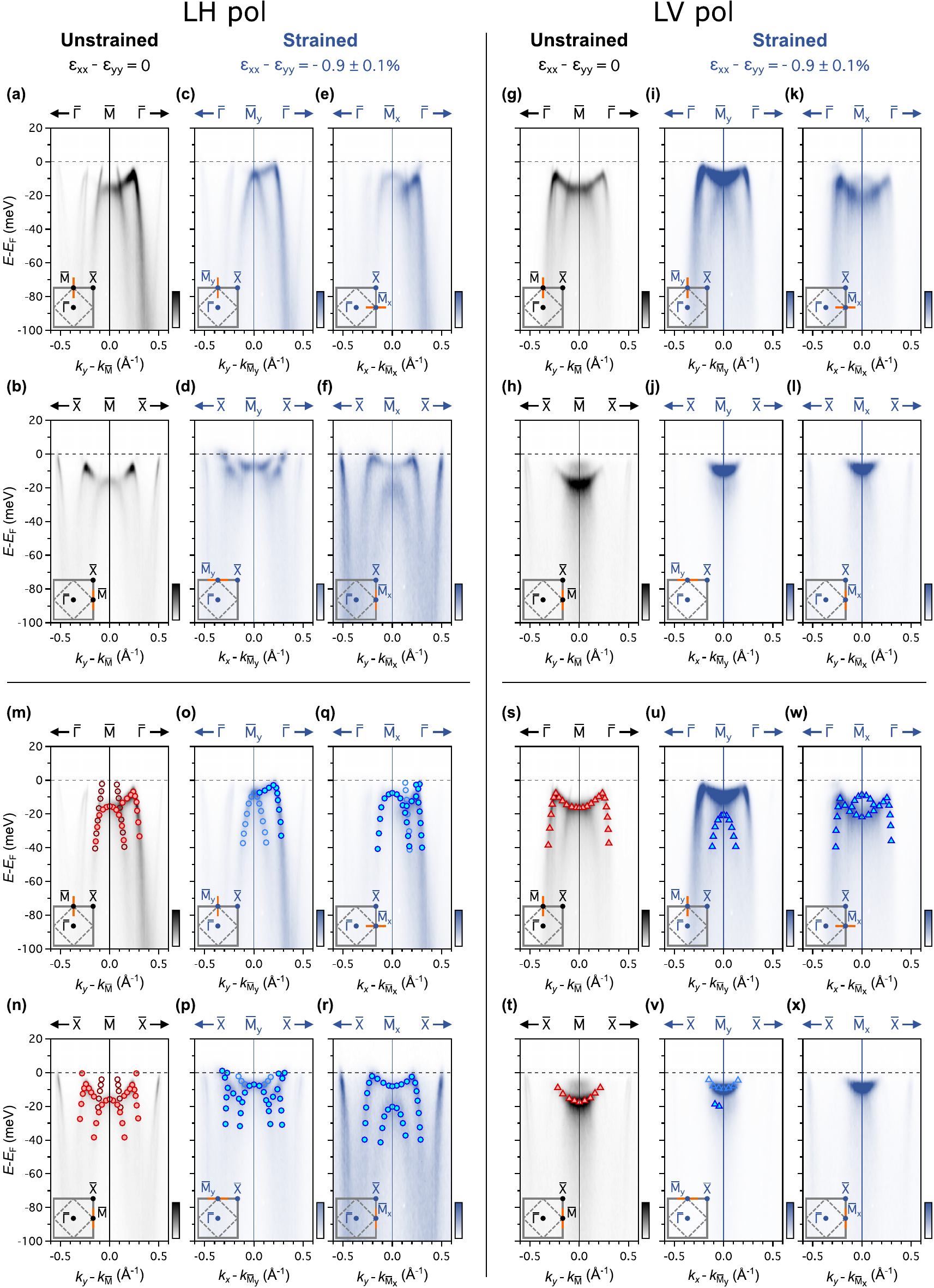}
\caption{\label{fig:figS7} (a-f) Single dispersions measured ($h\nu$=40eV) using LH-pol. (g-l) Equivalent dispersions measured using LV-pol. (m-x) Extracted band dispersions from the data shown in (a-l). The markers match the conventions in Fig. 3 of the main text and Supplementary Fig.~\ref{fig:figS8}. To improve readability, here we employ a different blue for the markers in the strained data.}
\end{figure}

\

\begin{figure}[!h]
\includegraphics{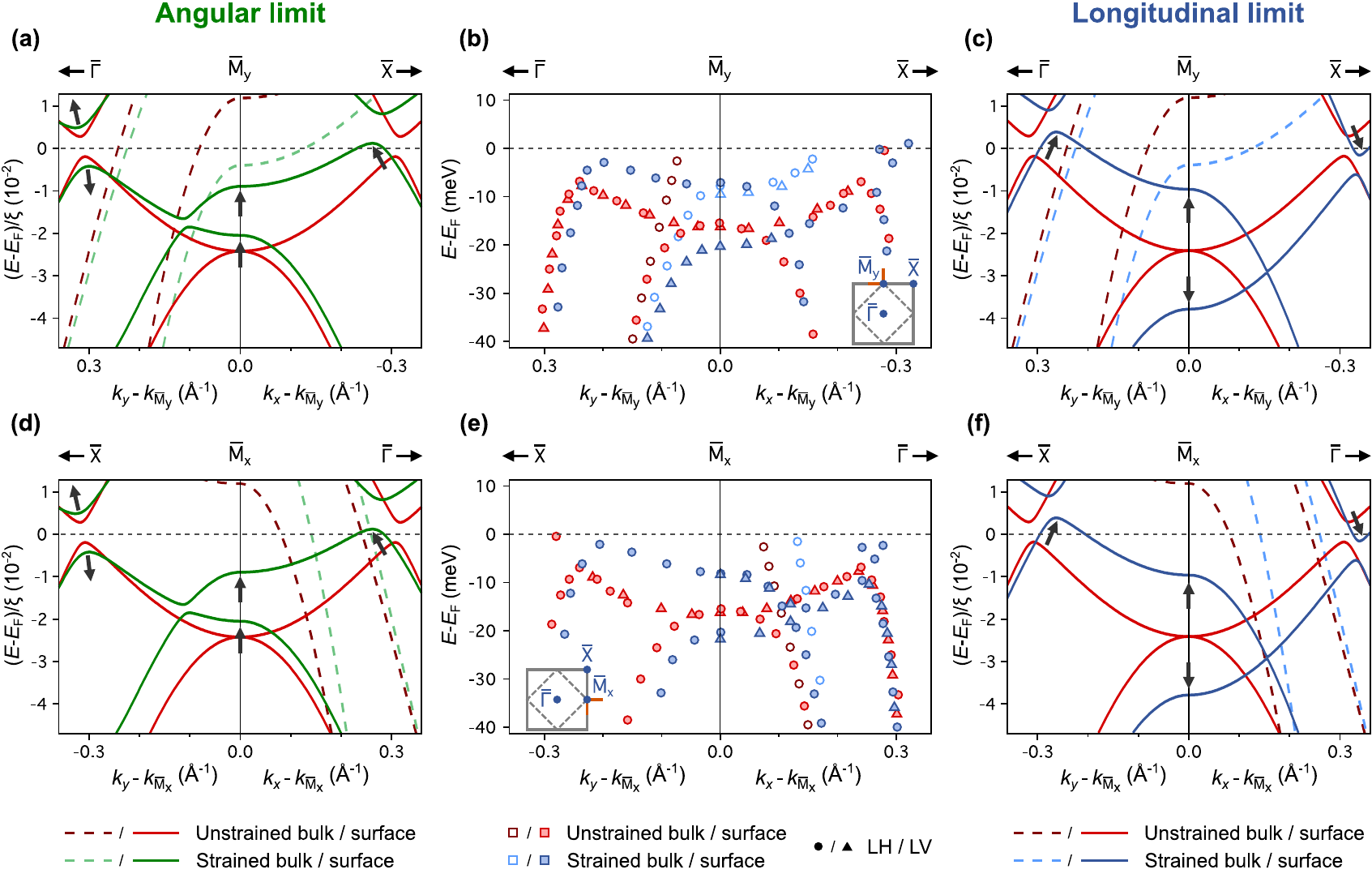}
\caption{\label{fig:figS8} (a-c) Angular (a) and longitudinal (c) limit tight-binding calculations of the surface electronic structure along the $\overline{\mathrm{\Gamma}}$-$\overline{\mathrm{M}}_y$-$\overline{\mathrm{X}}$ direction, and (b) corresponding extracted band dispersions from our ARPES data, reproduced from Fig. 3 of the main text. (d-f) Equivalent calculations and measurements for the $\overline{\mathrm{X}}$-$\overline{\mathrm{M}}_x$-$\overline{\mathrm{\Gamma}}$ direction. While the $\overline{\mathrm{\Gamma}}$-$\overline{\mathrm{M}}_y$-$\overline{\mathrm{X}}$ and $\overline{\mathrm{X}}$-$\overline{\mathrm{M}}_x$-$\overline{\mathrm{\Gamma}}$ directions are formally equivalent in the surface Brillouin zone, the ARPES matrix elements for measurements along these two directions vary markedly (Fig.~\ref{fig:figS7}). Nonetheless, the dispersions extracted from our data points are the same within experimental error. The bulk bands are also shown (dotted lines in the calculations, open symbols in the data).}
\end{figure}

\newpage

\section{Structural model}

\begin{figure}[h!]
\includegraphics{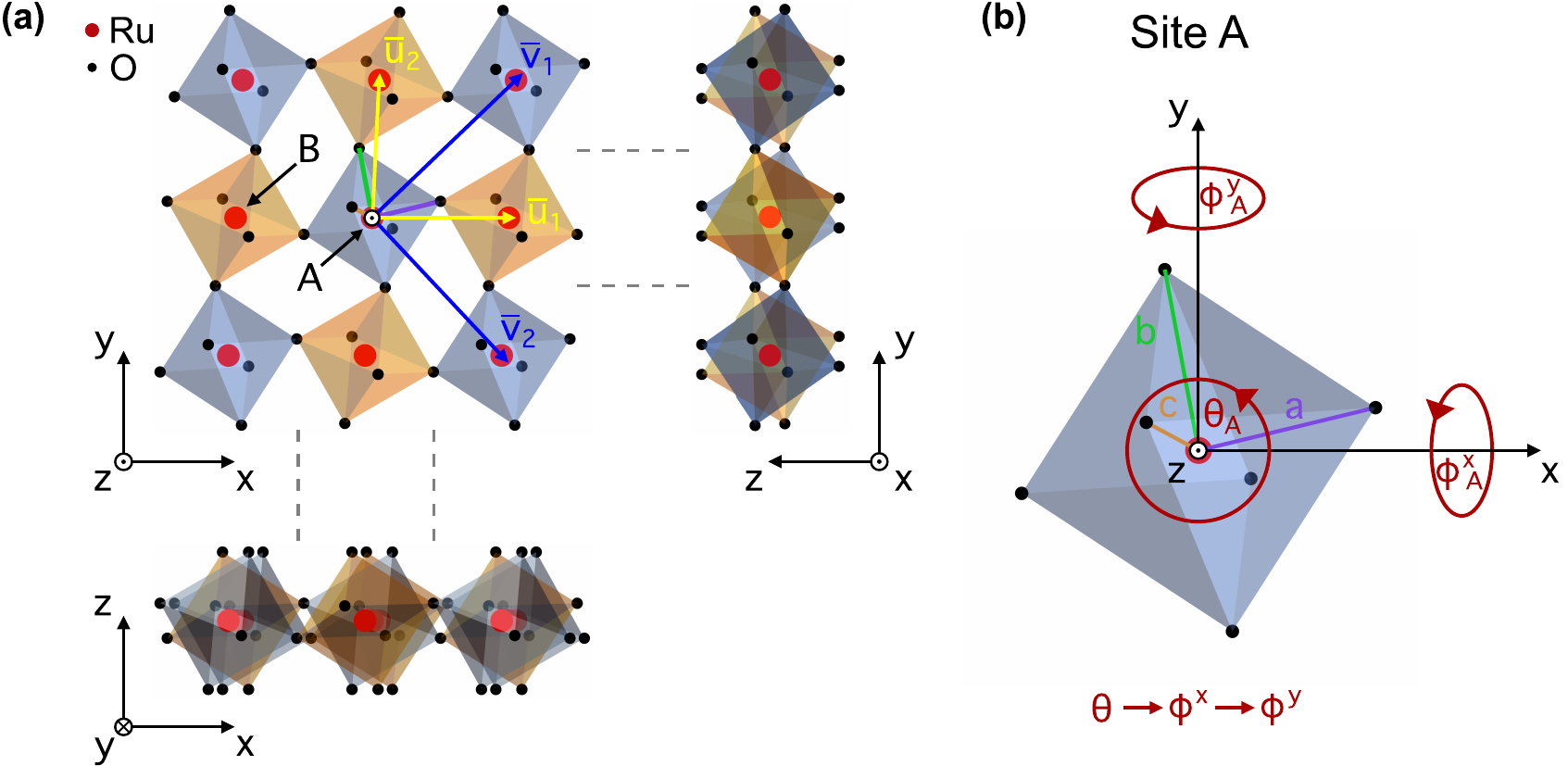}
\caption{\label{fig:figS9} (a) Bipartite rigid octahedra model in the monoclinic configuration ${(C_1=-0.23{\AA}^2,C_2=0.71{\AA}^2)}$ as seen from different projections. The rotation angles are given by ${(\theta_A=13.3\degree,\phi^x_A=-8.6\degree,\phi^y_A=-16.4\degree)}$ and ${(\theta_B=-13.7\degree,\phi^x_B=15.1\degree,\phi^y_B=10.7\degree)}$, while the octahedra lengths were chosen such that ${a=b=c}$. (b) The RuO$_6$ octahedron at the A-sites in (a). The octahedron lengths and rotation angles are defined with respect to the bulk Sr\textsubscript{2}RuO\textsubscript{4} coordinate system.}
\end{figure}

\begin{figure}[t]
\includegraphics{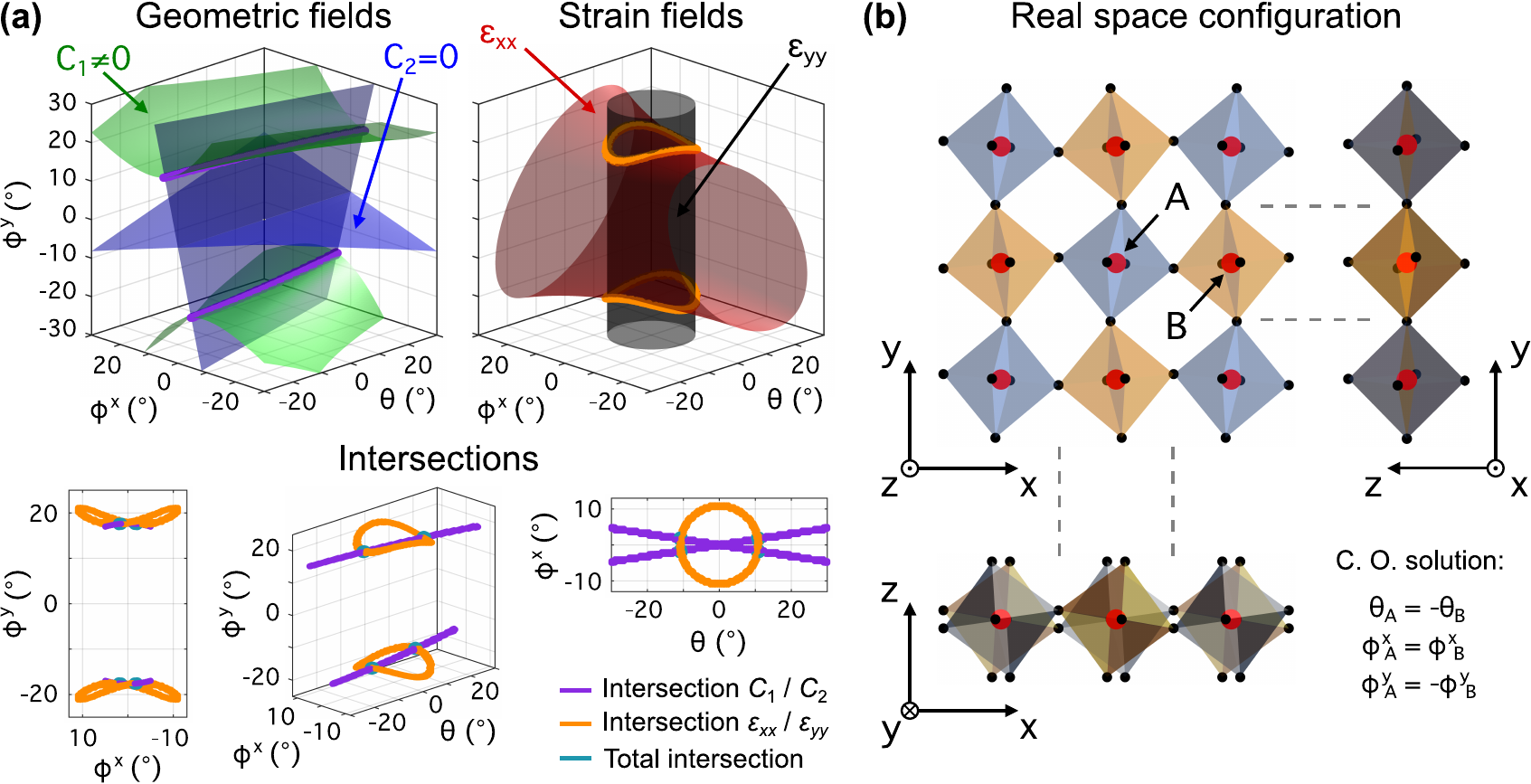}
\caption{\label{fig:figS10} (a) Geometric field isosurfaces ${(C_1=-0.36{\AA}^2,C_2=0{\AA}^2)}$, strain field isosurfaces ${(\varepsilon_{xx}=-0.9\%,\varepsilon_{yy}=0.3\%)}$ and their intersections along different projections for a centered orthorhombic system. As seen from symmetry, the total intersection is unique up to sign changes leading to equivalent configurations. (b) Real space configuration of the octahedra derived from the total intersection in (a). The centered orthorhombic solution obeys a particular set of rules constraining the possible octahedra rotations. The explicit rotation angles are given by ${(\theta_A=-5.4\degree,\phi^x_A=-0.8\degree,\phi^y_A=-8.8\degree)}$.}
\end{figure}

The surface layer of Sr$_2$RuO$_4$ is made up of a bipartite lattice of corner-sharing RuO$_6$ octahedra. Some approaches to describe the geometry and energetics of such octahedra are given in Refs. \cite{Qian2020,Filipov2017}, however, we show here a simple model which focuses on a direct mapping between globally applied uniaxial stress and local octahedral rotations and distortions. The most general version of this is shown in Fig. \ref{fig:figS9}. The RuO$_6$ octahedra at the A-sites are parametrized by the lengths ${(a,b,c)}$ describing their distortion and the angles ${(\theta_A,\phi^x_A,\phi^y_A)}$ describing their rotational configuration, specified in Fig. \ref{fig:figS9}(b). The octahedra at the B-sites are characterized by angles ${(\theta_B,\phi^x_B,\phi^y_B)}$ analogous to the ones at the A-sites and we assume that their lengths are also given by ${(a,b,c)}$. All rotation angles are measured with respect to the bulk RuO\textsubscript{2} layer in Fig. \ref{fig:figS2} and the rotation order follows the Tait-Bryan convention $\theta\rightarrow\phi^x\rightarrow\phi^y$ in Fig. \ref{fig:figS9}(b), which indicates that the rotation around the $z$-axis is performed first and the rotation around the $y$-axis is carried out last \cite{Berner2008}. A series of constitutive equations relate the different sites to guarantee that the octahedra remain linked through their vertices, while the global configuration of the system is governed by:
\begin{equation}
\abs{\vec{u}_1}^2-\abs{\vec{u}_2}^2=\vec{v}_1\cdot{\vec{v}_2}=C_1
\label{528}
\end{equation}
\begin{equation}
\frac{\abs{\vec{v}_1}^2-\abs{\vec{v}_2}^2}{4}=\vec{u}_1\cdot{\vec{u}_2}=C_2,
\label{529}
\end{equation}
where $C_1$ and $C_2$ are geometric scalar fields that depend on the distortions and rotations of the individual octahedra and the vectors ${(\vec{u}_1,\vec{u}_2,\vec{v}_1,\vec{v}_2)}$ are represented in Fig. \ref{fig:figS9}(a). We note that the octahedra remain symmetric, i.e. they are perfectly elongated octahedra that are then tilted and rotated. From equations (\ref{528}) and (\ref{529}), if $C_1=0$ and $C_2=0$ the system is tetragonal; if only $C_1=0$ it is rectangular orthorhombic; if only $C_2=0$ it is centered orthorhombic; and when both $C_1$ and $C_2$ are non-zero it becomes monoclinic. If we apply a uniaxial stress along the directions $\vec{u}_1$ and $\vec{u}_2$, we can define the corresponding anisotropic strain scalar fields as:
\begin{equation}
\varepsilon_{xx}=\frac{\abs{\vec{u}_1}}{\abs{\vec{u}^0_1}}-1
\label{501}
\end{equation}
\begin{equation}
\varepsilon_{yy}=\frac{\abs{\vec{u}_2}}{\abs{\vec{u}^0_2}}-1,
\label{502}
\end{equation}
where $\vec{u}_1$, $\vec{u}_2$ ($\vec{u}^0_1$, $\vec{u}^0_2$) are the vectors after (before) the deformation, respectively. It can be shown that:
\begin{equation}
\varepsilon_{xx}=l_{xx}+r_{xx}+l_{xx}r_{xx}
\label{50}
\end{equation}
\begin{equation}
\varepsilon_{yy}=l_{yy}+r_{yy}+l_{yy}r_{yy},
\label{51}
\end{equation}
where $r_{xx}$ and $r_{yy}$ are the contributions to the total strain due to rotations only and $l_{xx}$ and $l_{yy}$ represent the strain generated purely from octahedral distortions. Thus, the \textit{longitudinal} and \textit{angular} limits can be defined as the case when $r_{xx},r_{yy}=0$ and $l_{xx},l_{yy}=0$, respectively. In the unstrained Sr\textsubscript{2}RuO\textsubscript{4} surface layer we have $\abs{\vec{u}^0_1}=\abs{\vec{u}^0_2}\equiv{u_0}$, where we apply a surface reconstruction given by ${(\theta_A=-\theta_B=-7\degree,\phi^x_A=\phi^x_B=0\degree,\phi^y_A=\phi^y_B=0\degree)}$. Thus, rearranging equations (\ref{501}) and (\ref{502}) and using equation (\ref{528}) for the strained lattice, we obtain:
\begin{equation}
C_1=u^2_0\left[(\varepsilon_{xx}+1)^2-(\varepsilon_{yy}+1)^2\right].
\label{517}
\end{equation}
Therefore, the configuration of the system under uniaxial stress is fully determined by the scalar fields ${(C_1,C_2,\varepsilon_{xx},\varepsilon_{yy})}$, where only three of them are independent. We can visualize such scalar fields in the rotational space ${(\theta,\phi_x,\phi_y)}$ and constrain them to a particular lattice geometry and $B_{1g}$ strain. For example, the configuration $(C_2=0{\AA}^2,\varepsilon_{xx}=-0.9\%,\varepsilon_{yy}=0.3\%)$ in the \textit{angular limit} ($l_{xx},l_{yy}=0$) represents a centered orthorhombic system with $-0.9\%$ compression along the $x$-axis and $33\%$ Poisson expansion along the $y$-axis, where all the strain develops into octahedral rotations. If the intersection of these surfaces exist, the rotational state of the octahedra for the required conditions can be uniquely determined. In Fig.~\ref{fig:figS10}(a) we show that the intersection does exist in this case, and in Fig. \ref{fig:figS10}(b) we plot the real space configuration of the RuO$_6$ octahedra derived from it. The position, distortion and rotational configuration of the RuO$_6$ octahedra obtained from our model define the geometry onto which we perform the tight-binding calculation.

\section{Model parameters}

The model parameters for the unstrained bulk and surface electronic structure, and for the strain calculations, for bulk, and surface in the \textit{longitudinal} (L.L.) and \textit{angular} (A.L.) limits, are given in  Table~\ref{tab:table1} below. The angles for the relevant octahedral rotations are listed in Table~\ref{tab:table2}.

\begin{table*}[h!]
\caption{\label{tab:table1}Parameters employed in our tight-binding calculations.}
\begin{ruledtabular}
\begin{tabular}{cccccccccc}
& $Z_{Ru}$ & $Z_{O}$ & $c_{d_{xy}}$ & $c_{e_g}$ & $g$ & $l_{xx}$ & $l_{yy}$ & $r_{xx}$ & $r_{yy}$\\ \hline
Bulk Unstrained & $7.8$ & $5.9$ & $-0.005$ & $0.14$ & $0.005$ & $0$ & $0$ & $0$ & $0$ \\
Bulk Strained & $7.8$ & $5.9$ & $-0.005$ & $0.14$ & $0.005$ & $-0.018$ & $0.006$ & $0$ & $0$ \\
Surf. Unstrained & $8.6$ & $6.3$ & $-0.005$ & $0.09$ & $0.005$ & $0$ & $0$ & $0$ & $0$ \\
Surf. L.L. & $8.6$ & $6.3$ & $-0.005$ & $0.09$ & $0.005$ & $-0.018$ & $0.006$ & $0$ & $0$ \\
Surf. A.L. & $8.6$ & $6.3$ &$-0.005$ & $0.09$ & $0.005$ & $0$ & $0$ & $-0.018$ & $0.006$ \\
\end{tabular}
\end{ruledtabular}
\end{table*}

\begin{table*}[h!]
\caption{\label{tab:table2}Angles defining octahedra rotation in the unstrained, angular and longitudinal limits.}
\begin{ruledtabular}
\begin{tabular}{ccccccc}
 & $\theta_A$ & $\phi^x_A$ & $\phi^y_A$ & $\theta_B$ & $\phi^x_B$ & $\phi^y_B$ \\ \hline
Surf. Unstrained & $-7$° & $0$° & $0$° & $7$° & $0$° & $0$° \\
Ang. limit & $-3.0645$° & $-0.6810$° & $-12.5222$° & $3.0645$° & $-0.6810$° & $12.5222$° \\
Long. limit & $-7$° & $0$° & $0$° & $7$° & $0$° & $0$° \\
\end{tabular}
\end{ruledtabular}
\end{table*}

\

\newpage

\end{document}